\documentclass[a4paper,11pt]{article}
\usepackage{pos}

\def\Li2{\rm{Li}_2}

\def\bbe{\bar{\beta}}

\def\alf1{ {\alpha\over\pi} }

\title{Overview of theoretical precision of the luminosity at future electron-positron colliders}

\author*[a]{B.F.L. Ward}
\author[b]{S. Jadach}
\author[c]{W. Placzek}
\author[b]{M. Skrzypek}
\author[d]{S.A. Yost}
\affiliation[a]{Baylor University,\\
Waco, TX, USA}

\affiliation[b]{Institute of Nuclear Physics, Polish Academy of Sciences,\\
  Kraków, Poland}

\affiliation[c]{Institute of Applied Computer Science, Jagiellonian University,\\
Krakow, Poland}

\affiliation[d]{The Citadel,\\
Charleston, SC, USA}

\emailAdd{bfl\_ward@baylor.edu}
\emailAdd{Stanislaw.Jadach@cern.ch}
\emailAdd{Wieslaw.Placzek@uj.edu.pl}
\emailAdd{Maciej.Skrzypek@ifj.edu.pl}
\emailAdd{yosts1@citadel.edu}

\abstract{For both the FCC-ee and the ILC, to exploit properly the respective precision physics program, the theoretical precision tag on the respective luminosity will need to be improved from the analogs of the $ 0.054 \% (0.061\%)$ results at LEP at $M_Z$, where the former (latter) LEP result has (does not have) the pairs correction. At the FCC-ee at $M_Z$ one needs improvement to $0.01\%$, for example. We present an overview of the roads one may take to reach the required $0.01\%$ precision tag at the FCC-ee and of what the corresponding precision expectations would be for the FCC-ee$_{350}$, ILC$_{500}$, ILC$_{1000}$, and CLIC$_{3000}$ setups.}
\centerline{BU-HEPP-22-06, Nov., 2022}
\FullConference{%
  41st International Conference on High Energy physics - ICHEP2022\\
  6-13 July, 2022\\
  Bologna, Italy
}

\begin{document}
\maketitle

\section{Introduction}
The importance of a precision MC event generator for the luminosity process at the FCC-ee and other future colliders has been emphasized in Refs.~\cite{fccwksp2019,fccwksp2018,blondel-jnt-2019}. Compared to the situation at the time of LEP, there has already been substantial progress. In what follows, we give an overview of the expectations for the theoretical precision of the luminosity at future electron-positron colliders with the current situation in mind. \par
Specifically, we show in Table~\ref{tab:lep-update-2019} a summary of the progress to date relative to the time of LEP.
\begin{table}[ht!]
\centering
\begin{tabular}{|l|l|l|l|}
\hline
Type of correction~/~Error
    &  1999
        & Update 2019
\\ \hline 
(a) Photonic ${\cal O}(L_e\alpha^2 )$
    & 0.027\% ~\cite{Jadach:1999pf}
        & 0.027\%
\\ 
(b) Photonic ${\cal O}(L_e^3\alpha^3)$
    & 0.015\%~\cite{Jadach:1996ir}
        & 0.015\%
\\ 
(c) Vacuum polariz.
    &0.040\%~\cite{Burkhardt:1995tt,Eidelman:1995ny} 
        & 0.011\%~\cite{fjeger-fccwksp2019,pjnt-stj-2019}
\\ 
(d) Light pairs
    & 0.030\%~\cite{Jadach:1996ca}
        & 0.010\%~\cite{Montagna:1998vb,Montagna:1999eu}
\\ 
(e) $Z$ and $s$-channel $\gamma$ exchange
    &0.015\%~\cite{Jadach:1995hv,Arbuzov:1996eq}
        & 0.015\%
\\ 
(f) Up-down interference
    &0.0014\%~\cite{Jadach:1990zf}
        & 0.0014\%
\\ 
(g) Technical Precision& -- & (0.027)\%
\\ \hline  
Total
    & 0.061\%~\cite{bhlumi-precision:1998}
        & 0.037\%
\\ \hline  
\end{tabular}
\caption{\sf
Summary of the total (physical+technical) theoretical uncertainty with BHLUMI~\cite{bhlumi4:1996} for a typical
calorimetric LEP luminosity detector within the generic angular range
of $18$--$52$\,mrad.
Total error is summed in quadrature.
}
\label{tab:lep-update-2019}
\end{table}
We see that, in spite of there being no really dedicated effort, considerable progress has been made. \par
\section{Current Situation, Related to LEP and BHLUMI Upgrade}
In Ref.~\cite{Jadach:2018} we have given the detailed description of the steps one would take on the path to improving the theoretical precision of BHLUMI the desired 0.01\% 
precision tag needed for future colliders like the FCC-ee on the $Z^0$ resonance. For completeness, we recall these steps in Table~\ref{tab:lep2fcc} and we refer the reader to Ref.~\cite{Jadach:2018}
\begin{table}[ht!]
\centering
\begin{tabular}{|l|l|l|l|}
\hline
Type of correction~/~Error
        & Update 2018
                &  FCC-ee forecast
\\ \hline 
(a) Photonic $[{\cal O}\left(L_e\alpha^2 \right)]\; {\cal O}\left(L_e^2\alpha^3\right)$
        & 0.027\%
                &  $ 0.1 \times 10^{-4} $
\\ 
(b) Photonic $[{\cal O}\left(L_e^3\alpha^3\right)]\; {\cal O}\left(L_e^4\alpha^4\right)$
        & 0.015\%
                & $ 0.6 \times 10^{-5} $
\\
(c) Vacuum polariz.
        & 0.014\%~\cite{JegerlehnerCERN:2016}
                & $ 0.6 \times 10^{-4} $
\\
(d) Light pairs
        & 0.010\%~\cite{ON1,ON2}
                & $ 0.5 \times 10^{-4} $
\\
(e) $Z$ and $s$-channel $\gamma$ exchange
        & 0.090\%~\cite{BW13}
                & $ 0.1 \times 10^{-4} $
\\ 
(f) Up-down interference
    &0.009\%~\cite{Jadach:1990zf}
        & $ 0.1 \times 10^{-4} $
\\
(g) Technical Precision & (0.027)\% 
                & $ 0.1 \times 10^{-4} $
\\ \hline 
Total
        & 0.097\%
                & $ 1.0 \times 10^{-4} $
\\ \hline 
\end{tabular}
\caption{\sf
Anticipated total (physical+technical) theoretical uncertainty 
for a FCC-ee luminosity calorimetric detector with
the angular range being $64$--$86\,$mrad (narrow), near the $Z$ peak.
Description of photonic corrections in square brackets is related to 
the 2nd column.
The total error is summed in quadrature.
}
\label{tab:lep2fcc}
\end{table}
for their detailed discussion accordingly. Here, we note that, with sufficient research support\footnote{We have to be realistic that, given the far off start times for the future electron colliders, such research support may also be far off.}, the path to the 0.01\% precision tag for BHLUMI is indeed an opened one.\par
As we illustrated in the Introduction, progress on the theoretical effort to improve the luminosity theory precision at electron colliders is occurring even though there is no really dedicated effort in that direction. We note the recent results in Refs.~\cite{signer:2021,engel:2022} which feature the exact NNLO correction to the Bhabha scattering (the luminosity process at electron colliders) with next-to-soft stabilization. These results should viewed relative to the results in Refs.~\cite{Jadach:1999pf,bhlumi-precision:1998,bhlumi4:1996,bhlumi2:1992,Jadach:1995hy} which realize the exact ${\cal O}(\alpha^2 L)$ corrections with the exact ${\cal O}(\alpha^2 L^2)$ correction done with amplitude-based resummation via MC event generator methods. To be more precise, if we write the NNLO cross section as 
\begin{equation}
\sigma^{(2)} = (\frac{\alpha}{\pi})^2L^2\sigma^{(2)}_2+(\frac{\alpha}{\pi})^2L\sigma^{(2)}_1+(\frac{\alpha}{\pi})^2\sigma^{(2)}_0,
\end{equation} 
then the results in Refs.~\cite{signer:2021,engel:2022} realize $\sigma^{(2)}_i,\; i = 0,\;1,\;2,$ while those in Refs.~\cite{Jadach:1995hy,bhlumi-precision:1998,bhlumi2:1992,bhlumi4:1996,Jadach:1999pf} realize $\sigma^{(2)}_i,\; i = 1,\;2,$ exactly. We note that the relevant big logarithm is $L = \ln(|t|/m_e^2)$ in an obvious notation and that the constant term
$\sigma^{(2)}_0$ enters at the level of $(\frac{\alpha}{\pi})^2\cong 5.4\times 10^{-6}$. From the standpoint of progress on cross-checks for precision theory in the Bhabha process, the authors in Refs.\cite{signer:2021,engel:2022} have made a comparison with the results from BABAYAGA~\cite{Balossini:2006wc,CarloniCalame:2011ny,babayaga-2019}: for the $\phi$ factory type setup of $\sqrt{s}= 1020$ MeV, with $E_{min}= 408$ MeV, $20^\circ < \theta_{\pm} < 160^\circ$ ($\theta_{\pm}$ are the respective $e^\pm$ cms scattering angles and $E_{min}$ is their minimum energy.), and the accolinearity $\zeta$ cut $\zeta_{max} = 10^\circ$, the agreement is at the level $0.07\%$. This is two orders of magnitude larger than the level at which $\sigma^{(2)}_0$ enters as noted above and suggests that one of the two calcautions may have unknown technical errors, as both are supposed to be exact at NNLO.\par
The semi-soft approximation used in Refs.~\cite{signer:2021,engel:2022} using the notation therein can be isolated via 
\begin{equation}
\lim_{\xi\to 0}\xi^2{\cal M}^{(\ell)}_{n+1}={\cal E}{\cal M}^{(\ell)}_{n}+\xi \Delta{\cal M}^{(\ell)}_{n+1}+\ldots,
\label{eq2}
\end{equation}
where ${\cal M}^{(\ell)}_{n}$ is the $\ell-\text{loop}$ matrix element squared amplitude with n final state particles, ${\cal E}$ is the respective YFS~\cite{yfs:1961} exponent and $\xi=2E_\gamma/\sqrt{s}$ the scaled photon energy. The semi-soft approximation keeps the $1/\xi$ non-universal term determined by $\Delta{\cal M}^{(\ell)}_{n+1}$ in eq.(\ref{eq2}). This 
should be compared with the corresponding treatment of the analogous effects in Refs.~\cite{Jadach:1995hy,bhlumi-precision:1998,bhlumi2:1992,bhlumi4:1996,Jadach:1999pf}. When the exact ${\cal O}(\alpha^2 L)$ correction in latter references is implemented in BHLUMI, the $(\frac{\alpha}{\pi})^2$ term will be missing from $\bar\beta^{(2)}_{1U}$ and $\bar\beta^{(2)}_{1L}$ in the cross section
for the process $e^+(p_1)+ e^-( q_1) \rightarrow e^+(p_2) +e^-( q_2) +\gamma(k_1) + ... +\gamma(k_n)+ \gamma(k'_1) + ... + \gamma(k'_{n'})$ as presented in Ref.~\cite{Jadach:1995yfsmc} which we reproduce here for completeness:\\
\begin{align}
\sigma^{(r)}
& = \sum_{n=0}^\infty \sum_{n'=0}^\infty
  {1\over n!}{1\over n'!}
   \int {d^3 p_2\over p_2^0}
   \int {d^3 q_2\over q_2^0}\;
    \prod_{j=1}^n \;
       \int\limits_{k_j   \notin \Omega_U}
              {d^3k_j\over k^0_j}       \tilde{S}_p(k_j)\;
    \prod_{l=1}^{n'} \;
       \int\limits_{ k'_l \notin \Omega_L}
              {d^3{k'}_l\over {k'}^0_l} \tilde{S}_q(k'_l)
\notag\\
& \delta^{(4)}\bigg(
          p_1-p_2 +q_1-q_2 -\sum_{j=1}^n k_j -\sum_{l=1}^{n'} {k'}_l
            \bigg)\;
 e^{Y_p(\Omega_U) + Y_q(\Omega_L)}
\notag\\
& \Bigg\{
  \bbe^{(r)}_0
 +\sum_{j=1}^n
    {\bbe^{(r)}_{1U}(k_j) \over \tilde{S}_p(k_j)}
 +\sum_{l=1}^{n'}
    {\bbe^{(r)}_{1L}(k'_l) \over  \tilde{S}_q(k'_l)}
 +\sum_{n \geq j>k \geq 1}\;\
    {\bbe^{(r)}_{2UU}(k_j,k_k)
                       \over \tilde{S}_p(k_j)\tilde{S}_p(k_k)}
\notag\\
& \qquad\qquad
 +\sum_{n' \geq l>m \geq 1}\;\
    {\bbe^{(r)}_{2LL}(k_l,k_m)
                       \over \tilde{S}_q(k'_l)\tilde{S}_q(k'_m)}
 + \sum_{j=1}^n \sum_{l=1}^{n'}\;
    {\bbe^{(r)}_{2UL}(k_j,k'_l)
                       \over \tilde{S}_p(k_j)\tilde{S}_q(k'_l)}
          \Bigg\}.
\label{master}
\end{align}
where we refer the reader to Ref.~\cite{Jadach:1995yfsmc} for the definitions of the respective IR functions and constructs.
There is no semi-soft approximation in the approach in eq.(3).\par
\section{Current Situation, Higher Energies}
In Ref.~\cite{Jadach:2021henl} we have extended our analysis in Ref.~\cite{Jadach:2018} to include the expectations for the luminosity theory error for higher energy future electron colliders: FCCee$_{350}$, ILC$_{500}$, ILC$_{1000}$, and CLIC$_{3000}$, where the collider 
cms energy in GeV is indicated by its subscript. We see in Table 3 that the key variable is the geometric mean momentum transfer $\bar{t}\equiv\sqrt{t_{min}t_{max}}$ for the respective acceptances for the attendant cms energies using an obvious notation. 
\begin{figure}[h]
\centering
\includegraphics[width=0.9\textwidth]{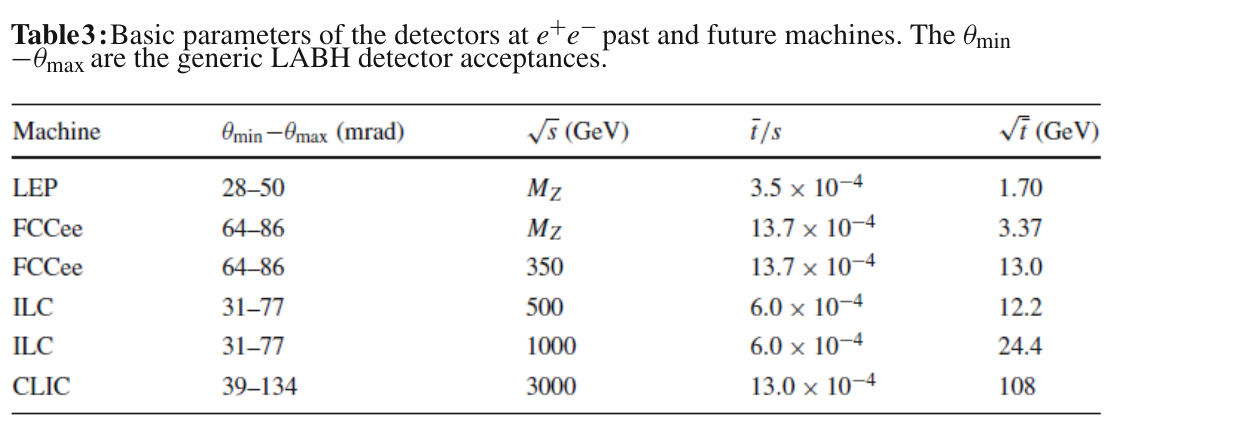}
\end{figure}
Generalizing~\cite{Jadach:2021henl} our methods first to the proposed 500 GeV ILC and subsequently to the FCCee$_{350}$, ILC$_{1000}$, and CLIC$_{3000}$, we get the analog of Table 2 for the ILC$_{500}$ which we show here in Table 4
\begin{figure}[h]
\centering
\includegraphics[width=0.9\textwidth]{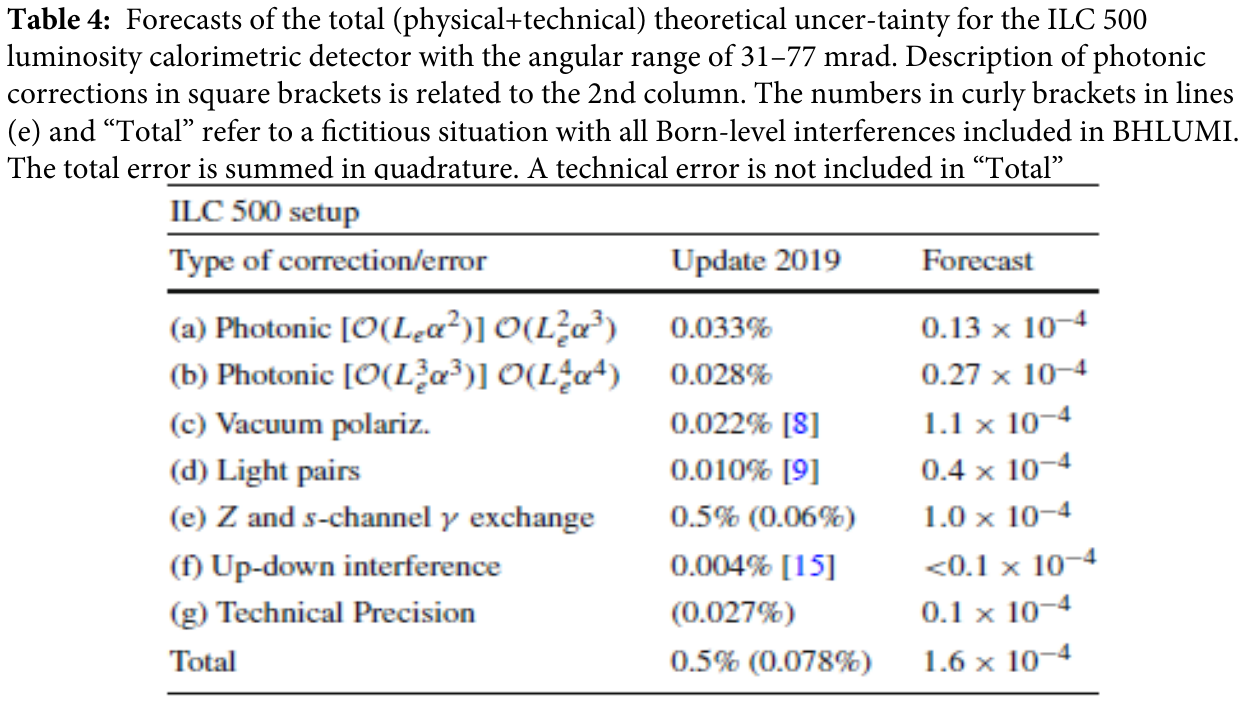}
\end{figure}
as well as forecasts for the FCCee$_{350}$, ILC$_{1000}$, and CLIC$_{3000}$ which we show here in Table 5. 
\begin{figure}[h]
\centering
\includegraphics[width=0.9\textwidth]{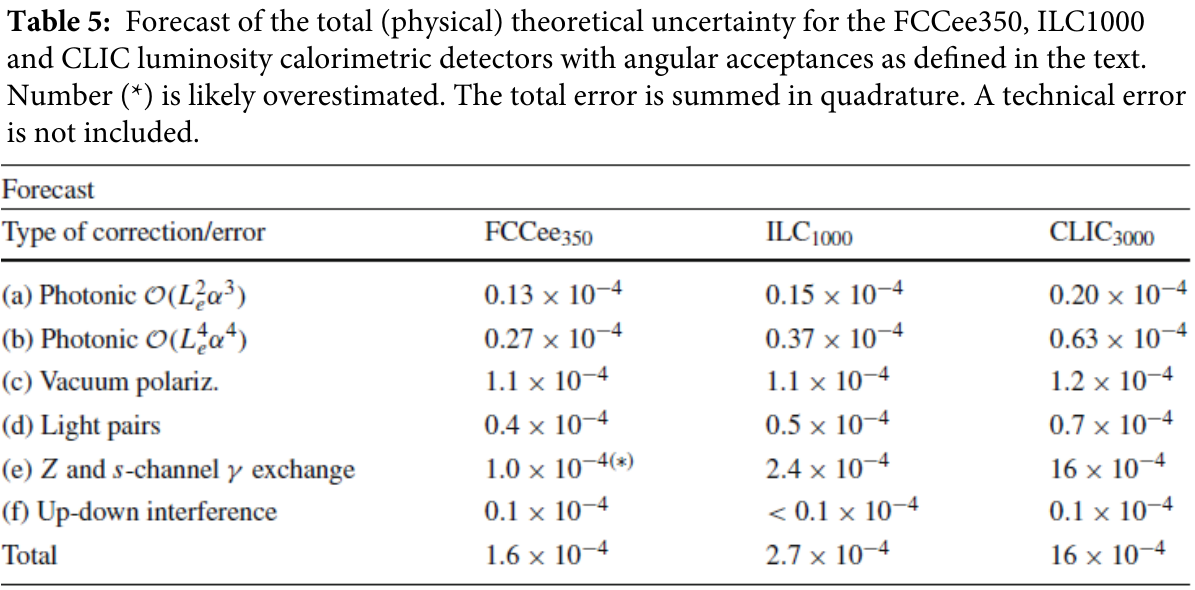}
\end{figure}
Of course, with appropriate resources, these results for the higher energies could all be improved as it is indicated in Ref.~\cite{Jadach:2021henl}.\par
We conclude with the following observations. The BHLUMI team, with appropriate resources, can improve it to meet the physics requirements of the planned future electron colliders. The far off nature of the colliders raises a legitimate question on the timing of these resources. It is a question that has yet to be answered.\par
\section*{Acknowledgments} This work was partially supported by the National Science Centre, Poland Grant No. 2019/34/E/\\
ST2/00457 and by The Citadel Foundation. We thank Dr. Patrick Janot for helpful discussion. Two of us (SJ and BFLW) thank Prof. Gian Giudice for the support and kind hospitality of the CERN TH Department. 
\setlength{\bibsep}{1.7pt}
\bibliography{Tauola_interface_design}{}
\bibliographystyle{utphys_spires}



\end{document}